\documentclass{emulateapj}

\shorttitle{The Rotation of sub-populations in $\omega$~Cen}
\shortauthors{Pancino et al.}

\begin{document}

\title{The Rotation of sub-populations in $\omega$~Centauri\footnote{Based on
data obtained with the Giraffe-FLAMES facility of ESO Very Large Telescope
during the Ital-FLAMES GTO programme 71.D-0217(A). Also based on data from the
VALD and GEISA databases.}}

\author{E. Pancino\footnote{e-mail: elena.pancino@oabo.inaf.it}}
\affil{INAF - Bologna Observatory, via Ranzani 1 I-40127 Bologna, Italy}

\author{A. Galfo and F.R. Ferraro}
\affil{Astronomy Department, Bologna University, via Ranzani 1 I-40127 Bologna, Italy}

\and

\author{M. Bellazzini}
\affil{INAF - Bologna Observatory, via Ranzani 1 I-40127 Bologna, Italy}

\begin{abstract} 

We present the first result of the Ital-FLAMES survey of red giant branch (RGB)
stars in $\omega$~Cen. Radial velocities with a precision of
$\sim$0.5~km~s$^{-1}$ are presented for 650 members of $\omega$~Cen observed with
FLAMES-Giraffe at the Very Large Telescope. We found that stars belonging to the
metal-poor (RGB-MP), metal-intemediate (RGB-MInt) and metal-rich (RGB-a)
sub-populations of $\omega$~Cen are all compatible with having the same
rotational pattern. Our results appear to contradict past findings by Norris et
al., who could not detect any rotational signature for metal-rich stars. The
slightly higher precision of the present measurements and the much larger sample
size, especially for the metal-richer stars, appear as the most likely
explanation for this discrepancy. The result presented here weakens the body of
evidence in favour of a merger event in the past history of $\omega$~Cen. 

\end{abstract}


\section{Introduction}

The giant globular cluster (GC) $\omega$~Centauri is one of the most studied
objects in the Milky Way since the 60s, thanks to its many peculiar properties.
As its main anomaly, $\omega$~Cen shows a wide abundance spread ($\sim$1~dex)
not only in the light elements, but also in the iron-peak elements
\citep{n95,n96,sk96,s00}, pointing towards a chemical evolution history more
reminiscent of a dwarf galaxy than a genuine GC. Moreover, its structural,
kinematical and dynamical properties appear atypical for a Galactic GC, having
properties more similar to Dwarf Elliptical galactic nuclei \citep{mv05}, and
being also an elongated \citep{g83,p03}, rotating system \citep{mer97} that is
not completely relaxed dynamically \citep{jay,f06}. 

Among the most recent, puzzling results we list the detection of complex
substructures in all evolutionary sequences, including the red giant branch
\citep[RGB,][]{l99,p00,s05a}, the subgiant branch and turn-off region
\citep[SGB,][]{f04,s05b} and the main sequence \citep[MS,][]{b04,p05}. These
studies have raised new questions that need to be answered, such as the debated
age differences among sub-populations \citep{hk00,hw00,s05b}, the possibility
that some sub-populations possess an extremely high Helium content
\citep{n04,p05} and others.

More related with the present paper, we point out a very interesting study on
the correlations between chemical and kinematical properties of $\omega$~Cen
carried out by \cite{n97}, where stars in the cluster were divided in two groups
based on their Ca abundance. They showed that metal-poorer stars rotate as the
majority of the cluster does, while metal-richer stars do not show any sign of
rotation. They also showed that the radial velocity dispersion appears to
decrease with metallicity, a very strange behaviour since metal-richer stars
tend to be concentrated to the cluster center, which is dynamically hotter.

\section{Observations and Data Treatment}
\label{pop}

\begin{figure}
\epsscale{1.22}
\plotone{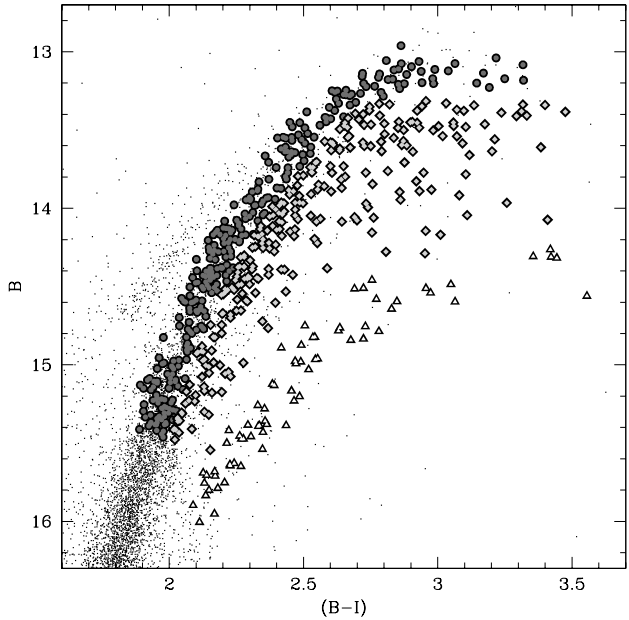}
\caption{Wide Field CMD of $\omega$~Cen from \citet{p00}. Stars observed with
Giraffe are highlighted: RGB-MP stars with dark grey circles, RGB-MInt with
light grey diamonds and RGB-a stars with white triangles.}
\label{cmd}
\end{figure}

A sample of $\sim$700 red giants was selected from the wide field photometry
originally published by \citet{p00} and \citet{tesi}, to derive accurate
chemical abundances of stars spanning the whole metallicity range of
$\omega$~Cen. Special care was taken to include a large fraction of
high-metallicity stars, which are the less studied up to now. All program stars
are distributed within 15~arcmin from the cluster center. Figure~\ref{cmd} shows
the program stars, marked on the B, (B--I) color magnitude diagram (CMD) of
\citet{p00}. Following the classification scheme of \citet{p00} and \citet{p03},
stars are divided in three groups: the metal-poor population (RGB-MP) comprising
75\% of the cluster giants, with [Fe/H]$\sim$--1.6; the metal-intermediate
population (RGB-MInt) comprising 20\% of the cluster giants, with
[Fe/H]$\sim$--1.2 and the metal-rich or anomalous stars (RGB-a), 5\% of the
cluster giants, with [Fe/H]$\simeq$--0.6 \citep{p02}. Sample sizes are shown in
Table~\ref{medie}.

Observations were done with FLAMES \citep{flames} at the ESO VLT in Paranal,
Chile, between the 22nd and 28th of May 2003, within the Guaranteed Observing
Time of the Ital-FLAMES Consortium. FLAMES was used in Medusa combined mode,
feeding 8 fibers to UVES and 132 fibers to Giraffe. To ensure the maximum
homogeneity in data quality and treatment, here we present the first results
obtained with Giraffe only, using the high-resolution (R$\simeq$22500) grating
HR13 (6120--6395\AA) and reaching a S/N$\simeq$50-100 per pixel, depending on
the star magnitude. 

Data were reduced with the Giraffe BLDRS (Base-Line Data Reduction
Software)\footnote{\tt http://girbldrs.sourceforge.net}, that includes cosmic
rays removal, bias subtraction, flat field correction, wavelength calibration
and pixel resampling. The version of the pipeline we used does not include
inter-order background and sky subtraction, but these have no significant effect
on radial velocity determinations.

Radial velocities were determined using DAOSPEC\footnote{\tt
http://cadcwww.hia.nrc.ca/stetson/daospec/;
\\ http://www.bo.astro.it/~pancino/projects/daospec.html} (Stetson \&
Pancino, in preparation), a program that automatically measures
equivalent widths of absorption lines in high-resolution spectra. Radial
velocities are determined by cross-correlation of the measured line
centers and a set of laboratory wavelengths. We used 150 clean and
unblended lines of the most common species obtained from the
VALD\footnote{\tt http://www.astro.uu.se/$\sim$vald/} \citep[Vienna
Atomic Line Database,][]{vald}. Observed radial velocities are the
average of the velocity obtained for each line after a 3$\sigma$
clipping, and the associated uncertainty is $\sigma /\sqrt n$, where $n$
is the number of lines used. Typical uncertainties are of the order of
0.13~km~s$^{-1}$.

\begin{table}
\caption{Radial Velocity Measurements}
\begin{tiny}
\begin{tabular}{lcccccc}
\tableline \tableline
WFI\tablenotemark{a} & RA\tablenotemark{b} & Dec\tablenotemark{b} & 
B\tablenotemark{a} & I$^{a}$ & V$_r$ & Pop\tablenotemark{c} \\
--- & (deg) & (deg) & (mag) & (mag) & (km~s$^{-1}$) & 
--- \\
\tableline 
100981 & 201.81686 & -47.59584 & 13.20 & 10.48 & 247.377$\pm$0.226 & 1 \\
102141 & 201.97864 & -47.58834 & 14.51 & 12.36 & 237.910$\pm$0.382 & 1 \\ 
102242 & 201.84398 & -47.58797 & 14.53 & 12.35 & 258.894$\pm$0.455 & 1 \\ 
103979 & 201.82367 & -47.57775 & 14.43 & 12.25 & 234.816$\pm$0.197 & 1 \\ 
104113 & 201.83088 & -47.57697 & 15.35 & 13.32 & 239.967$\pm$0.781 & 2 \\ 
\tableline  \tableline
\end{tabular}
\end{tiny}
\tablecomments{Table \ref{dati} is published in its entirety in the 
electronic edition of the {\it Astrophysical Journal Letters}.  A portion is 
shown here for guidance regarding its form and content.}
\tablenotetext{a}{WFI star number and B, I magnitudes from \citet{p00}.}
\tablenotetext{b}{Coordinates obtained using the astrometric catalog by
\citet{v00}.}
\tablenotetext{c}{Population classification according to \citet{p00} and
\citet{p03}; 1 stands for RGB-MP, 2 for RGB-MInt and 3 for RGB-a (see text for
details).}
\label{dati}
\end{table}

Since radial velocities were not the main goal of the observing programme,
simultaneous calibration lamps were not taken during the observations. We used
laboratory wavelengths of the telluric lines of the 6300\AA\  O$_2$ absorption
band from the GEISA\footnote{\tt
http://ara.lmd.polytechnique.fr/htdocs-public/products\-/GEISA/HTML-GEISA/}
database \citep[Gestion et Etude des Informations Spectroscopique
Atmospheriques,][]{geisa} to find a common zeropoint to our observed velocities.
The typical correction is less than $\pm$1~km~s$^{-1}$, with an uncertainty of
$\sim$0.45~km~s$^{-1}$, but we found that for some exposures there were larger
offsets, up to 3~km~s$^{-1}$. Finally, we applied the heliocentric correction
using the {\em rvcorrect} task within IRAF\footnote{Image Reduction and Analysis
Facility. IRAF is distributed by the National Optical Astronomy Observatories,
which is operated by the association of Universities for Research in Astronomy,
Inc., under contract with the National Science Foundation.}. The resulting
heliocentric radial velocities have a typical uncertainty of 0.5~km~s$^{-1}$ and
are reported in Table~\ref{dati}. 

Contaminating field stars were easily eliminated since their average velocity is
--4.0$\pm$38.5~km~s$^{-1}$, in very good agreement with Galactic model
predictions for Disk stars \citep{disk} and very different from the typical
velocity of $\omega$~Cen. After removing all stars with V$_r\leq$190~km~s$^{-1}$,
the final sample contains 649 cluster members, with an average V$_r$=233.4 and
$\sigma_r$=13.2~km~s$^{-1}$. Star by star comparisons with other catalogues yield
a very good agreement. For instance, the average radial velocity difference of
the 136 stars in common with \citet{may97} is
$\Delta$V$_r$=0.4$\pm$1.4~km~s$^{-1}$, while for the 53 stars in common with
\citet{sk96} it is $\Delta$V$_r$=--0.5$\pm$1.8~km~s$^{-1}$ and for the 382 stars
in common with \citet{rei06} it is $\Delta$V$_r$=--2.4$\pm$4.6~km~s$^{-1}$.

\section{Results}
\label{res}

We used the photometric definition of populations described in
Section~\ref{pop}, together with our radial velocity measurements, to construct
rotation curves for sub-populations in $\omega$~Centauri. To this aim, we first
de-projected RA and Dec into $X'$ and $Y'$ coordinates with the following
relations \citep[see also][]{v06}, suited for extended objects that are not
close to the celestial equator: 

\begin{displaymath}
X' = -r_0 \cos \delta \sin (\alpha - \alpha_0)
\end{displaymath}
\begin{displaymath}
Y' = r_0 (\sin \delta \cos \delta_0 - \cos \delta \sin \delta_0 \cos
(\alpha - \alpha_0))
\end{displaymath}

\noindent where $r_0$=10800/$\pi$ is the scale factor to have X' and Y' in
arcminutes. 

\begin{figure}
\epsscale{1.42}
\plotone{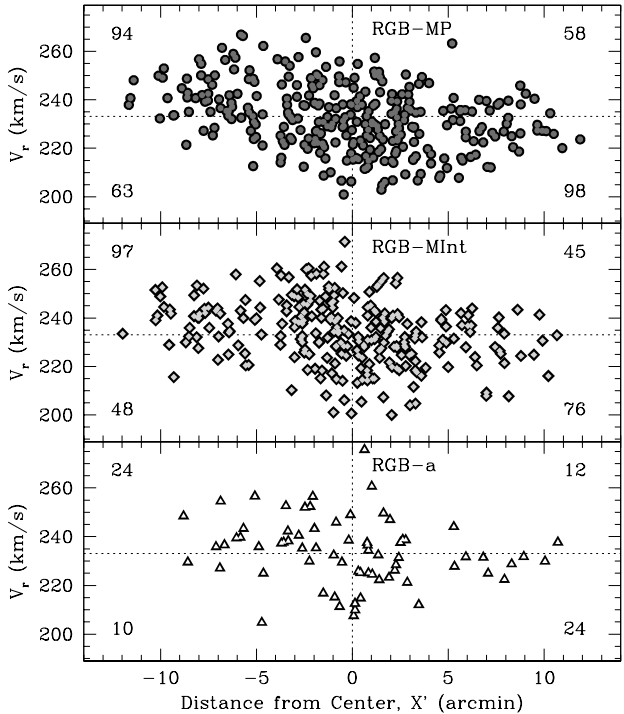}
\caption{Rotation curve for the RGB-MP (top panel), RGB-MInt (middle panel) and
RGB-a (bottom panel) samples. The typical rotation pattern of $\omega$~Cen can
be clearly seen not only for the RGB-MP population, but also for the RGB-MInt
and the RGB-a. Each panel is divided in four quadrants by dotted lines passing
through the cluster center and the cluster average radial velocity. Each
quadrant is labelled with the number of stars it contains.}
\label{curva}
\end{figure}

As a second step, we searched for the orientation of the rotation axis $\theta$
that maximizes the amplitude of the rotation signal. We found a relatively broad
maximum around $\theta=0$ and thus, for simplicity, we adopt a rotation axis
aligned with the North-South direction, which corresponds to the Y' axis and to
the minor isophotal axis \citep{g83,p03}. We therefore plotted radial velocities
of each sub-population against $X'$ (Figure~\ref{curva}) and found a clear
signature of rotation not only for the RGB-MP, as was expected, but also for the
RGB-MInt and the RGB-a. The numbers of stars in the four quadrants of each
rotation curve show that the rotational signature is strong even for the most
sparse population, i.e., the RGB-a. Assuming that the rotation curves are
symmetrical around the minor axis, we estimated the rotation velocity V$_{rot}$
near the maximum \citep[as found by][]{mer97,v06} as the average of V$_r$
between $6'$ and $8'$ from the center (see Table~\ref{medie}). As can be seen,
the values of V$_{rot}$ are comparable to each other, within the uncertainties,
and are compatible with previous literature estimates.

If the three samples were non-rotating, the histogram of radial velocities on
the Eastern half of the cluster would be identical to the Western half (see
Figure~2). A Kolmogorov-Smirnov test shows instead that the radial velocity
distribution of stars on the Eastern side of the cluster has a very low
probability of being extracted from the same parent distribution of the Western
side. More quantitatively, the probabilities P$_{KS}$ derived from the
cumulative distributions are: 10$^{-5}$ for the RGB-MP, 5$\times$10$^{-6}$ for
the RGB-MInt and 5$\times$10$^{-3}$ for the RGB-a.

\subsection{Literature comparison} 

A correlation between kinematic properties and chemical abundances in
$\omega$~Cen has been attempted, to our knowledge, only by \citet{n97} and
\citet{s05b}. \citet{n97} correlated the Calcium IR triplet abundances by
\citet{n96} with the radial velocity measurements by \citet{may97}, for a
global sample of $\sim$400 RGB stars. The sample was split in a metal-poor
group with [Ca/H]$\leq$--1.2 and a metal-rich group with [Ca/H]$>$--1.2. The
metal-poor group contains mainly stars that we classify as RGB-MP, with some
contamination by RGB-MInt stars, while the metal-rich group contains mainly
RGB-MInt and a handful of RGB-a stars, with very little contamination by RGB-MP
stars.

\begin{figure}
\epsscale{1.42}
\plotone{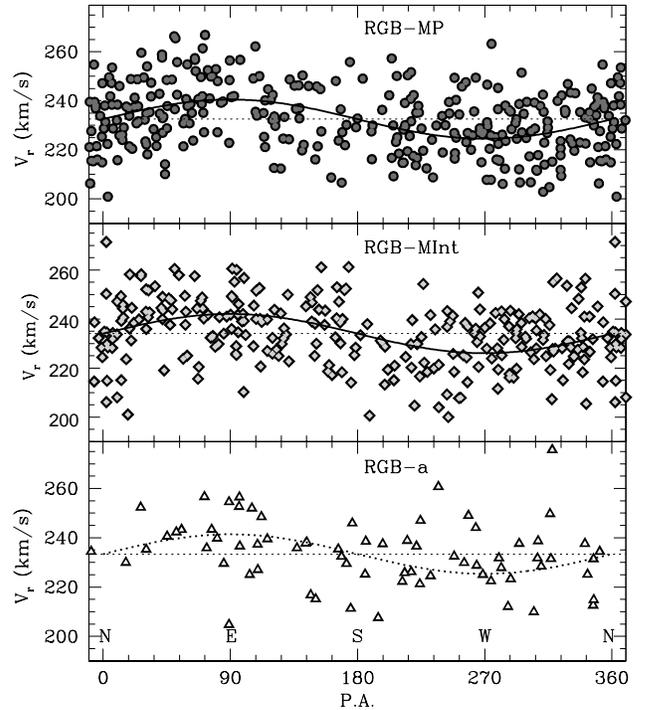}
\caption{Radial velocity measurements are plotted against the position angle
(P.A., counted from North towards East) for the  RGB-MP (top panel), the RGB-MInt
(middle panel) and the RGB-a (bottom panel). Sinusoids with an amplitude of
$\sim$8~km~s$^{-1}$ \citep{mer97} are overplotted for reference.}
\label{noi}
\end{figure}

The two main results presented by \citet{n97} are {\em (i)} metal-poor stars
rotate, while the metal-rich population does not show any sign of rotation (at
a 2~$\sigma$ level of confidence) and {\em (ii)} the radial velocity dispersion
decreases as metallicity increases, i.e. the metal-rich population is
kinematically cooler than the metal-poor one, even if it is concentrated in the
central, hotter regions of the cluster\footnote{This was referred to as the
radial velocity dipersion paradox in $\omega$~Cen, and a possible explanation,
involving the presence of a face-on metal-rich disk, was presented by
\citet{v99}. Some support to this hypothesis comes from the possible presence
of a small disc-like structure in the center of $\omega$~Cen pointed out by
\citet{v06}.}. A correlation between radial velocity dispersion and metallicity
was also reported by \citet{s05b}, who found a similar trend as in \citet{n97},
except that the dispersion increases again for the most metal rich stars in
their sample.

\begin{table}[t]
\caption{Sample sizes, radial velocity and rotation velocity.}
\begin{tiny}
\begin{tabular}{lrcccccc}
\tableline\tableline
Population & $n$ & V$_r$         & $\sigma_r$    & A$_{rot}$     & 
V$_{rot}~(6'$--$8')$ & $n~(6'$--$8')$ \\
---        & --- & $(km/s)$ & $(km/s)$ & $(km/s)$ & $(km/s)$     & --- \\
\tableline
Whole        & 649 & 233.4 & 13.2 & $\sim$6 & 6.8$\pm$1.0 & 87 \\
RGB-MP       & 313 & 232.5 & 13.3 & $\sim$7 & 8.1$\pm$1.5 & 46 \\
RGB-MInt     & 266 & 234.2 & 13.1 & $\sim$6 & 5.3$\pm$1.7 & 33 \\
RGB-a        &  70 & 234.0 & 13.4 & $\sim$4 & 6.0$\pm$3.0 &  8 \\
\tableline
\end{tabular}
\end{tiny}
\label{medie}
\end{table}

Velocity dispersions may be quite sensitive to outliers and to the radial
distribution of the chosen samples, and a detailed analysis would greatly benefit
from the metallicity estimates of individual stars that are not yet available.
Here we simply note that there appears to be no difference in the global velocity
dispersions of the three subsamples (Table~\ref{medie}). However, the present
photometric classification is probably too coarse to reveal subtle effects such
as the ones presented by \citet{n97} and \citet{s05b}, hence we postpone any
conclusion on radial velocity dispersions to a future paper. On the other hand,
strong coordinate motions like the rotation patterns shown in Figure~\ref{curva}
are clearly very robust to the effect of a few outliers and can be profitably
studied with the available information. Even in the absence of individual
metallicity estimates, Figure~\ref{curva} leaves very little room for a
non-rotating sub-population within $\omega$~Cen. To facilitate a comparison with
Figure~3 by \citet{n97}, Figure~\ref{noi} shows the run of V$_r$ with the
position angle (P.A.) for the three populations of $\omega$~Cen. As before, all
populations show the well known rotation along the East-West direction
\citep{n97,mer97}. Even the RGB-a, despite the smaller sample, is still
compatible with the same rotational pattern.  The amplitude A$_{rot}$ of the
rotation signal in Figure~\ref{noi}, obtained with a $\chi^2$ minimization, is
reported in Table~\ref{medie}. However, A$_{rot}$ can only be considered as an
indicative value since it contains contributions from stars at very different
radii, and the systemic rotation is highly radius dependent. 

We considered two possible reasons why the rotational signature\footnote{As
stated before, the data presented in this paper do not allow to reach a firm
conclusion on the velocity dispersion, so we will only consider the rotation
patterns in the following discussion.} was not found before: sample size and
measurement precision. The size of the entire sample is not so different:
\citet{n97} had $\sim$400 stars, we have $\sim$650. However our sample contains
more metal-rich stars: we have 313 stars in the RGB-MP, and 336 stars in the
RGB-MInt and RGB-a together, while \citet{n97} had $\sim$300 metal-poor stars
and less than 100 metal-rich stars. Also, the precision of the radial velocity
measurements could play a role, although we have an uncertainty of
$\sim$0.5~km~s$^{-1}$ and \citet{may97} have an uncertainty which is only
slightly higher and still below 1~km~s$^{-1}$. However, as \citet{n97} made
clear, the absence of rotation for the metal-rich group was confirmed only at
the 2~$\sigma$ level, therefore the combination of a slightly higher precision
and a much larger sample size are probably enough to explain why we were able
to reveal such a signature with our dataset.

\section{Conclusions}

We presented the first results from the Ital-FLAMES survey of the RGB of
$\omega$~Cen. Radial velocities with uncertainties of 0.5~km~s$^{-1}$ are derived
for 650 radial velocity members of $\omega$~Cen, and are in very good agreement
with previous literature measurements. The main result obtained is that all the
three sub-populations of $\omega$~Cen show the same rotation pattern. These
findings appear in contradiction with the results presented by \citet{n97}. We
show that a combination of higher precision in the V$_r$ measurements and of a
larger sample size, especially for metal-rich stars (RGB-MInt and RGB-a), is the
most likely cause for the discrepancy. 

The results by \citet{n97} were the main piece of evidence in support of a
merger event in the past evolution of $\omega$~Cen. The evidence presented here
suggests that there is no rotational anomaly in $\omega$~Centauri. The only
other remaining evidence, apart from the radial velocity dispersion paradox
mentioned above, that still points towards a complicated dynamical history is
the structural difference among populations found by \citet{p03} and the
somewhat debated differential proper motion for the RGB-a found by \citet{f02},
questioned by \citet{pl03} but supported by \citet{hw04}. Clearly, a deeper
investigation into these aspects is now needed to finally settle the issue. 

\acknowledgments

We thank P.~B. Stetson, A.~Sollima and C.~Cacciari. We also thank J.~E.~Norris,
the referee of this paper, for his constructive comments. The financial support
from the PRIN-INAF 2005, PRIN-INAF2006 and ASI-INAF I/023/05/0 grants is
acknowledged.

\end{document}